\documentclass[12pt]{article}

\usepackage{scicite}

\usepackage{times}

\usepackage{amsmath}
\usepackage{amsfonts}
\usepackage{amssymb}
\usepackage{graphicx}
\usepackage[labelfont=bf, labelsep=period]{caption}

\usepackage{soul} % for the command \hl
\usepackage{color}
\usepackage{pdfpages}

\newenvironment{sciabstract}{%
\begin{quote} \bf}
{\end{quote}}

\title{Breaking Abbe's diffraction limit with harmonic deactivation microscopy}

\author
{Kevin Murzyn$^{1\dagger}$, Maarten L. S. van der Geest$^{1}$, Leo Guery,$^{1}$ Zhonghui Nie$^{1}$,\\ Pieter van Essen$^{1}$, Stefan Witte$^{1,2}$,  Peter M. Kraus$^{1,2\ast}$\\
\\
\normalsize{$^{1}$Advanced Research Center for Nanolithography (ARCNL)}\\
\normalsize{Science Park 110, 1098 XG Amsterdam, Netherlands}\\
\normalsize{$^{2}$Department of Physics and Astronomy, and LaserLaB, Vrije Universiteit}\\
\normalsize{De Boelelaan 1081, 1081HV Amsterdam, Netherlands}\\
\\
\normalsize{To whom correspondence should be addressed; E-mail: $^\dagger$ k.murzyn@arcnl.nl,  $^\ast$ p.kraus@arcnl.nl.}
}

\date{}

\begin{document} 


\baselineskip24pt


\maketitle


% The abstract should be a single paragraph, not to exceed 250 words and ideally closer to 200, written in plain language that a general reader can understand. It should include
% An opening sentence that states the question/problem addressed by the research AND
% Enough background content to give context to the study AND
% A brief statement of primary results AND
% A short concluding sentence.
% Do not include citations or undefined abbreviations in the abstract. Any abbreviations that appear in the title should be defined in the abstract.

\begin{sciabstract}
Nonlinear optical microscopy provides elegant means for label-free imaging of biological samples and condensed matter systems. The widespread areas of application could even be increased if resolution was improved, which is currently limited by the famous Abbe diffraction limit. Super-resolution techniques can break the diffraction limit but rely on fluorescent labeling. This makes them incompatible with (sub-)femtosecond temporal resolution and applications that demand the absence of labeling. 
Here, we introduce harmonic deactivation microscopy (HADES) for breaking the diffraction limit in non-fluorescent samples. By controlling the harmonic generation process on the quantum level with a second donut-shaped pulse, we confine the third harmonic generation to three times below the original focus size and use this pulse for scanning microscopy. We demonstrate that resolution improvement by deactivation is more efficient for higher harmonic orders, and only limited by the maximum applicable deactivation-pulse fluence. This provides a route towards sub-100~nm resolution in a regular nonlinear microscope. The new capability of label-free super-resolution can find immediate applications in condensed matter physics, semiconductor metrology, and biomedical imaging. %This last sentence can go out of abstract an in outlook

\end{sciabstract}

% In setting up this template for *Science Advances* papers, both
% the \section* command and the \paragraph* command are used for topical
% divisions.  Which you use will of course depend on the type of paper
% you're writing.  Review Articles tend to have displayed headings, for
% which \section* is more appropriate; Research Articles, when they have
% formal topical divisions at all, tend to signal them with bold text
% that runs into the paragraph, for which \paragraph* is the right
% choice.  Either way, use the asterisk (*) modifier, as shown, to

\section*{Introduction}
Far-field optical microscopy has impacted the fields of medicine, biology, and biophysics \cite{Weisenburger2015} like kaum any other technique. Microscopy is used as a daily diagnostics tool as well as equipment for cutting-edge scientific advancements. Microscopy benefits from its compactness, ease of use, and the possibility to image 2D as well as 3D objects\cite{Mertz2019}. 
Third harmonic generation (THG) is used as a label-free microscopy  \cite{Duncan1982} method to study cells \cite{Debarre2006,Witte2011}, photoresists \cite{Kallioniemi2020} and 2D materials\cite{Zhou2020}. For THG microscopy, an excitation pulse is focused into a material and generates harmonics. The high penetration depth of the near-infrared (NIR) excitation light and the sensitivity to changes in chemical composition are central advantages of THG compared to other optical contrast mechanisms. One challenge preventing an even more widespread application is the limited resolution. The reported resolution of THG microscopes is approximately 1 µm or slightly less \cite{Yildirim2015,Murakami2022}, which is far from the 200-nm resolution limit commonly stated in a confocal fluorescence microscope\cite{Shao2011}. This apparent limitation stems from the fact that in typical confocal scanning microscopes, the resolution is proportional to the focus size of the excitation laser of the emission process. This laser is typically in the infrared for harmonic microscopy but in the visible or UV range for fluorescent microscopy. To improve this resolution limit of harmonic microscopy, the emission has to be confined below the diffraction limit. Here we show that the unique optical properties of high-harmonic genreation from condensed matter enable sub-diffraction harmonic emission and thus super-resolution microscopy.
%This requires near-unity-contrast all-optical emission control – a standout property of high-harmonic generation, as we discuss below.
%Improving resolution and ultimately breaking the diffraction limit requires near-unity all-optical emission control – a standout property of high-harmonic generation.
\\
High-harmonic generation (HHG) is a laser-driven frequency upconversion process \cite{mcpherson87a,Ferray1988}, that converts an ultra-short laser pulse into a shorter-wavelength pulse at integer multiples of the fundamental carrier frequency. The discovery of solid–state HHG in 2010 \cite{Ghimire2011} also rejuvenated the interest in harmonic generation of lower orders below the bandgap \cite{juergens19a} and drove the development of a stringent microscopic theoretical framework \cite{golde08a,schubert14a} that complements the phenomenological framework of perturbative non-linear optics. On the quantum level, HHG is governed by laser-driven electron dynamics and contains signatures of the dynamic electronic structure of the generation medium that can be atoms \cite{shiner11a}, molecules \cite{Smirnova2009,kraus15b} and solids \cite{Ghimire2011,ghimire19a}. Inversely, slight variations to these electronic dynamics can have a large impact on the efficiency of HHG \cite{Kraus2013}. In a variety of solids, reversible deactivation of high-harmonics followed the photoexcitation of charge carriers\cite{Wang2017,Heide2022,Wang2022,Bionta2021,Cheng2020,Geest2023}. Different physical processes are considered to contribute to the deactivation \cite{VanEssen2024} such as state-blocking and ground-state depletion, a decrease in decoherence time \cite{Cheng2020,Geest2023,Brown2022}, and control of electron-hole recollisions in shaped driving waveforms\cite{Wang2023}. The modulation depth, or contrast, achieved can be close to 100\%, demonstrating near-unity-contrast emission control. This intensity modulation is the key enabler of super-resolution HHG microscopy by spatially confining HHG with an optical control pulse.\\ 
We took inspiration from the super-resolution techniques in fluorescence microscopy, which can provide resolutions below 10 nm \cite{Weber2022}. These techniques can be divided into two groups. The first group uses wide-field illumination of the sample and achieves super-resolution through stochastic on-and-off switching of fluorescence combined with localization algorithms \cite{Betzig1995,Betzig2006,Rust2006,Sharonov2006}. The second group consists of deterministic methods that rely on targeted modulation or saturation of fluorescence \cite{Hell1994,Klar2000,Hell1995,Gustafsson2005}. %Previous authors reduced the resolution of linear transmission microscopy in silicon \cite{Pinhas2018} and graphene \cite{Wang2013} based on the experimental principle of stimulated emission depletion (STED). 
Here, we utilize the high modulation depth of HHG \cite{Nie2023} for an approach similar to the one used in stimulated emission depletion (STED) microscopy \cite{Klar2000} for HHG microscopy. We label this new method as harmonic deactivation microscopy (HADES). In the present manuscript, we focus on THG, but the method will work similarly and with even better resolution for higher harmonic orders. In HADES, we add a donut-shaped deactivation pulse to the infrared driving pulse of HHG. This donut pulse deactivates harmonic emission in the outer parts of the point-spread function (PSF) of the excitation pulse leading to a narrower PSF compared to THG. A similar methodology has been shown \cite{Pinhas2018,Wang2013} for linear microscopy and patented \cite{Kanou2017} for coherent anti-Raman scattering microscopy.\\
The combination of ultrafast temporal and nanoscale resolution will be a unique advantage of HADES: Due to the necessity of intramolecular vibrational energy redistribution before stimulated emission in STED, fluorescence super-resolution is inherently incompatible with femtosecond time resolution \cite{penwell2017}. \\
While deactivation is a general phenomenon, we chose NbO$_2$ as a sample in the present manuscript. NbO$_2$ is a strongly correlated material that undergoes a non-thermal ultrafast insulator-to-metal phase transition (IMT) upon photoexcitation. This phase transition makes strongly correlated materials interesting for a new class of electronic devices such as memristors \cite{yang2011,kumar2020}. In recent work HHG was used to probe this transition in NbO$_2$ and measure its transition time scale to be $\sim$ 100 fs \cite{Nie2023}. The ability to add super-resolution to resolving IMTs will enable resolving the emergence of new phases with femtosecond temporal and nanometer spatial resolution. %While deactivation is a general phenomenon, NbO$_2$ is a proven candidate with a near 100\% HHG suppression following photoexcitation, and interesting by itself due to the ultrafast nature of the phase transition. %% commented out since it's kinda repetitive.

\section*{Results}
%Three different detector set-ups can detect the harmonic emission generated (dashed boxes inside fig. 2a). In short, the spectrometer resolves the harmonics emitted in the visible and UV range into the spectral components, the PSF imaging part reimages the PSF of the third harmonic and donut pulse used and the point scanning detector detects each shot for fast laser scanning microscopy. We describe the detectors in more detail in conjunction with their respective results. The detectors work for transmitted harmonics as well as for back-reflected harmonics. In reflection, the harmonics are separated from back-reflected light from the modulation pulse and NIR pulse by a dichroic mirror and a glass wedge, respectively.
\paragraph*{Deactivation of third harmonic generation.} 
The concept of the experiment is illustrated in Fig. 1.
In regular harmonic microscopy, a NIR excitation laser generates harmonic multiples of its fundamental frequency. 
For a Gaussian intensity distribution I$_{NIR}$ of the fundamental NIR excitation laser in focus (Fig. 1a), the intensity distribution of the third harmonic will also be Gaussian (Fig. 1c). The spatial emission intensity distribution, called point-spread function (PSF), of HHG is narrower than for the excitation pulse and scales in the perturbative regime with I$_{HHG}\propto$ I$_{NIR}^n$, where $n$ is the power law exponent of the process which is approximately three for THG. In this work, we further deactivate HHG spatially by adding a pulse that is donut-shaped in focus to HHG (Fig. 1b). This donut-shaped pulse is achieved by imparting an orbital angular momentum onto the deactivation pulse by means of a spiral phase plate inserted in the far-field pulse profile. This concept allows deactivating HHG to a narrower emission peak below the diffraction limit (Fig. 1d), since HHG is deactivated in regions of non-zero intensity of the donut-shaped deactivation pulse. By narrowing the spot size below the diffraction limit, we can implement a super-resolution label-free scanning microscope.\\
We use a modified Mach-Zehnder interferometer to realize these experiments (Fig. S1). The fundamental HHG driving pulse is centered at a wavelength of 1830 nm, while the donut pulse is centered at 400 nm. Both pulses are linearly and parallel polarized. Detailed descriptions of all experiments can be found in the materials and methods section. \\ 
We first measure the spectra of the emitted harmonics. We vary the delay between the pulses (Fig. 2a) as well as the intensity of the deactivation pulse (Fig. 2b). The spectral shape of the third harmonic (inset Fig. 2b) stays identical for all fluences. However, with increasing modulation fluence the amplitude and the integrated signal reduce. We normalize the integrated signal to its deactivation, i.e. a normalized deactivation of one means that we measure no signal, while zero means the signal is unchanged. When the driving pulse arrives before the deactivation pulse no change in the THG can be observed. This is followed by a sharp increase in the maximum deactivation when both pulses arrive simultaneously. At increasing time delays the signal recovers with a fast and slow exponential component, consistent with the results in \cite{Nie2023}.\\
The deactivation becomes stronger the higher the deactivation-pulse fluence. At the highest deactivation-pulse fluence of 24 mJ/cm$^2$, the integrated signal amounts to 2$\%$ of the reference at the temporal overlap of the two pulses. From this fluence dependence of the normalized deactivation, we determine a saturation fluence $\Phi_{sat}$, defined as the fluence at half deactivation, of 2.5 mJ/cm$^2$, while the intensity modulation depth is 98\%. Comparing these results to deactivation curves for the fifth harmonic (grey curve Fig. 2b) shows that higher-order harmonics have a lower saturation fluence of 1.8 mJ/cm$^2$.\\
%% Original Version: The suppression can be attributed to two processes. First, the waveform of the two overlapping but orthogonally polarized pulses will deflect electron-hole recollisions at temporal overlap (i.e. at and around a delay of zero) analogous to HHG driven by multicolor fields in gases \cite{abbing2020,roscam2021}. Second, for later times (i.e. for delays larger than zero) electron scattering is increased. This is because an excited carrier fraction increases scattering and reduces the dephasing time \cite{becker88a}, and also because the ultrafast phase transition creates a metallic phase, where electron scattering is generally larger. Scattering and the associated dephasing-time reduction act as amplitude suppression \cite{Nie2023} window onto the time-domain emission of HHG, thus reducing the overall emission intensity.
%% New version:
Suppression of HHG can be attributed to two processes. First, two-color waveforms strongly modulate HHG \cite{abbing2020,roscam2021}, and the waveform of two overlapping electric fields (i.e at a time delay of zero) with incommensurate frequencies can decrease the overall intensity within a harmonic order \cite{Negro2011}. Second, right after excitation and for later times (i.e. for delays of and larger than zero) electron scattering is increased. On the one hand, an excited carrier fraction increases scattering and reduces the dephasing time \cite{Heide2022,Geest2023}, i.e. the time until which electron-hole pairs recombine coherently \cite{becker88a}. On the other hand, excitation of NbO$_2$ triggers an ultrafast phase transition on a time scale of 100~fs \cite{Nie2023} into a metallic phase, where electron scattering is generally larger. Scattering and the associated dephasing-time reduction act as amplitude suppression\cite{Nie2023} window onto the time-domain emission of HHG, thus reducing the overall emission intensity. Both effects - waveform modulation and increased electron dephasing are contributing to deactivation in our experiment. This is evidenced by the observation of highest levels of HHG deactivation at time zero for all pump fluences, which can thus be partially attributed to waveform modulation. Furthermore, the persisting high level of HHG deactivation after time zero can be attributed to increased scattering due to excited electrons and also an insulator-to-metal phase transition at higher pump fluences. At a time delay of zero, where the experiments described in this article are performed, both effects can thus play a role.

\paragraph*{Point-spread function reduction.} We first estimate the expected resolution improvement of HADES numerically. We combine the fluence-dependent deactivation with the spatial distribution of the intensity of the donut pulse and multiply this with an initial Gaussian distribution of the emitted harmonics. Assuming a maximum fluence of 25 mJ/cm$^2$, this numerical model results in a PSF reduction by a factor of three (Fig. S2). This requires that the donut has a constant azimuthal fluence distribution and that the maximum of the donut coincides with the FWHM of the THG. We compare the result to a modified formula for the resolution limit of STED microscopy \cite{Westphal2005}:
\begin{equation}
    d = \frac{\lambda_0}{2\text{NA}\sqrt{n}}\frac{1}{\sqrt{1+\zeta}}.
\end{equation}
The first term containing the wavelength of the fundamental excitation laser $\lambda_0$, nonlinear order $n$ (three or five for detection of the third or fifth harmonic, respectively, in the perturbative regime) and numerical aperture NA, is the diffraction limit after Abbe modified for harmonic microscopy. The second term gives the resolution improvement based on the saturation level $\zeta = \Phi^{\text{peak}}_{\text{HADES}}/\Phi_{sat}$. If we assume a peak fluence $\Phi^{\text{peak}}_{\text{HADES}}$ of 24 mJ/cm$^2$ and the measured saturation fluence for the third harmonic the resolution improves by a factor of 3.3, which is close to our numerical simulations. For fifth harmonic generation, we calculate a resolution improvement of 3.8. This shows pathways to improve the resolution even further by decreasing the saturation fluence, e.g. via optimizing the deactivation scheme or by using higher-order harmonics that deactivate at lower fluences.\\
We now demonstrate this all-optical control and shrinkage of the PSF. A lens with a focal length of 25 cm generates the third harmonic in the sample (Fig. 3a). A home-built bright-field microscope images the PSF in transmission. We calculate the diffraction limit of the lens that focuses the driving pulses as 40.7$\pm$1.2 µm, which is the first term in eq. (1). The unchanged third harmonic PSF (Fig. 3b) has an FWHM of 40$\pm$1.5 µm, which is in agreement with the theoretical diffraction limit. When the donut pulse (inset figure \ref{fig:PSF}c) deactivates the third harmonic emission the area that emits light is drastically reduced to an FWHM of 14$\pm$0.6 µm, while the peak intensity stays similar. This yields a reduction of factor 2.9$\pm$0.15, which matches the prediction of the numerical model. HADES’s spot therefore is roughly three times smaller than Abbe’s diffraction limit in eq. (1). This paves the way towards a feasible approach for label-free super-resolution scanning microscopy through HADES, which we demonstrate in the next section in a proof-of-principle experiment.
\paragraph*{Microscopic imaging.} For rapid sampling, the fundamental pulse of 800 nm was used since a similar deactivation behavior to 400 nm was observed. We used a lens with a focal length of 5 cm as the focusing objective, which also captured the back-reflected THG for imaging (Fig. 4a). In the detection pathway, a series of bandpass filters (BP) select a specific harmonic order. A tube lens then focuses the filtered light onto an avalanche photodiode (APD). The APD measures signal with low average powers on a shot-to-shot basis, enabling rapid scanning of the sample. A boxcar integrator filters out noise outside the 2\% duty cycle. Figures 4b) and c) show the recorded pictures for THG microscopy and HADES, respectively. These pictures depict the transition from an area with laser-induced damage to an undamaged sample area. In comparison, the HADES picture visualizes smaller details compared to THG microscopy. The super-resolution becomes visible when convolving the HADES image with a Gaussian. This fully reproduces the THG microscopy image (Fig. S3), whch demonstrates the resolution improvemnet by HADES. The two profiles (Fig. 4d) quantify this resolution improvement further. Along the dotted line in the THG microscope, a single peak is imaged while by using HADES two peaks become distinguishable. These two peaks are separated by 9.7 µm. The dotted line depicts a single peak at the very edge of the thin film. With THG microscopy the FWHM of this emitter is 15.6 µm, while, with HADES the FWHM reduces to 7.1 µm. Notably, the two profiles are close to perpendicular to each other. We show, therefore, that the resolution improves in two dimensions at the same time.
\section*{Discussion}
We now turn to discussing signal formation in HADES. The image in Fig. 4 shows an area of the NbO$_2$ thin film that was thermally damaged prior to the HADES experiments. In bright-field microscopy, the dark area is visible as well, indicating that material was ablated (Fig. S4a) when damage occured. The bright spot (dashed profile in Fig. 4) is located at a pointy tip of the thin film to pure substrate transition seen in dark-field microscopy (Fig. S4b). A combination of multiple effects can be the reason for the enhancement THG signal enhancement at htis bright spot. First, the third-order susceptibility of the sample can be changed locally, since the NbO$_2$ thin film generates harmonics more efficiently than the pure substrate. Second, near-field confinements of the electric field of the fundamental in subwavelength nanostructures can enhance the HHG efficiency \cite{RoscamAbbing2022}. Indeed, the presence of such sub-wavelength nanostructures becomes visible under a scanning electron microscope (Fig. S5). During experimenting, we generally observed an enhancement of the signal when scanning over an edge in the sample, indicating near-field enhancement of the fundamental field, which boosts the efficiency of HHG locally. \\
Our current experimental procedure was challenged by the available low repetition rate (2~kHz) / high-pulse energy (3.5~mJ) laser system. The high pulse energies required a strong attenuation of both the modulation and excitation pulse by effective optical densities up to 5, which made detectable signals weak due to the 2-kHz repetition rate. These requirements made optimization more challenging especially for shorter focal lengths lenses. Furthermore, scanning microscopy as shown in Fig. 4 will benefit from either high-speed scanning stages or ideally a galvo-mirror system that scans the position of the focus through the objective, neither of which was available for the current experiment. Implementing both - higher repetition rate lasers and high-speed scanning will make taking microscopy images in seconds possible. This will also allow focusing the HHG driving pulses with high-NA objectives instead of lenses (Fig. 4). Since we have demonstrated the mechanism for breaking the diffraction limit in harmonic microscopy in this manuscript, we now understand that we can quantitatively predict the resolution improvement by means of Eq.~1 and the harmonic deactivation curve (Fig. 2b). \\
The in-depth analysis of the deactivation process measured here and by \cite{Nie2023} delivers a three-time resolution improvement of THG microscopy and a 3.8-time resolution improvement for a fifth harmonic generation microscope. In conjunction with a near-infrared excitation at 1800 nm and a high NA objective of 1.35, the expected resolutions for the specific set of wavelengths, polarizations, and material (NbO$_2$) chosen here are 120 nm for THG and 80 nm for the fifth harmonic generation. Deepening the understanding of the deactivation process for harmonic generation in materials will lead to an even better resolution by either reducing the saturation fluence, i.e. finding ways to deactivate HHG more efficiently, or by extending the highest applicable peak fluence. This principle is ultimately not limited to the deactivation of HHG by photoexcited carriers, which limits the maximum applicable fluence due to photo-induced damage. Other deactivation methods are also suited for this, i.e. control of the electron-hole recombination with a combination of below band-gap driving wavelengths might also be a suitable pathway \cite{Wang2023}. Ultimately, resolution in HADES is only limited by the real-space excursion of electrons and holes by the driving laser pulse. This excursion underlies the microscopic currents that radiate HHG, and corresponds to a real-space displacement on the single-digit nanometer scale, the exact value of which depends on electronic structure and laser parameters. As a hypothetical deactivation below this limit would actually truncate HHG even from the inner region without a deactivation pulse, this excursion length scale sets the ultimate resolution limit for HADES. 
Furthermore, other optical nonlinearities such as coherent anti-Stokes Raman scattering \cite{Beeker2011}, stimulated Raman scattering \cite{Zhang2010}, or sum frequency generation \cite{Sekiguchi2008} can be used to achieve super-resolution with similar concepts. However, the extreme nonlinearity of HHG creates a maximum sensitivity to control pulses. This likely makes HHG and related high-order frequency mixing processes the ideal candidates for resolution improvement by deactivation in particular, and optical emission control and optical switching in general.

%In conclusion, we show full optical control over the resolution of a third-harmonic generation (THG) microscope. This can be achieved by suppressing the third harmonic using a NIR laser to increase the excited carrier population. Further, we show that the PSF can be reduced to half the limit predicted by the Abbe criterion. This reduced PSF can then be used to improve the resolution of a scanning THG microscope by a factor of two. Optimizing the focal spot sizes of the two laser pulses will lead to a further improvement in resolution. With this technique called harmonic deactivation microscopy (HADES), we observed new features close to a damaged spot of an NbO$_2$ thin film on top of a silicon substrate. 

\section*{Materials and Methods}
\subsection*{Materials}
The sample investigated in this study is the polycrystalline thin film NbO2, which was grown on a c-plane sapphire substrate through reactive bias target ion pulse deposition (RBTIBD), and the more information of the material synthesis are available in previous literature\cite{wang2015,Nie2023}. The Raman spectrum and X-ray diffraction 2$\theta$ scan are both utilized to confirm the high quality and the bct lattice structure of our sample. Any other oxide components (NbO or Nb2O5) have not been found. The thickness of our sample is $\sim$115 nm, measured by X-ray reflectivity, and its surface roughness is less than 1 nm. Moreover, the sapphire substrates are double-side polished, ensuring the possibility of transmission measurements.

\subsection*{Methods}
\paragraph*{Laser system and experimental setup.}
 A Ti:Sapphire laser amplifier (Solstice, Spectra-Physics) produces a laser pulse at a central wavelength of 790 nm and with a pulse duration below 72 fs. The repetition rate is set to 2 kHz while the average output power is 7 W. A pulse splitter (BS) sends the laser light into each of the arms. In the excitation arm, the light pumps an optical parametric amplifier (OPA, Topas prime), which emits NIR light centered at 1800 nm. The idler of the parametric process was used to generate the third harmonic since only the idler can reach the desired wavelength. In the modulation arm, the delay of the 800 nm pulse can be adjusted by a mechanical delay stage. This is called the modulating arm. Afterwards, a vortex phase plate (PP) converts the intensity profile into a donut shape. Optionally, a second harmonic generation BBO can be introduced after the vortex phase plate to change the central wavelength to 400 nm. Variable attenuators control the average power in each arm independently, while telescopes set the pulse size. A dichroic mirror (DM) combines the two pulses. After this, an objective lens (Obj) focuses this pulse bundle onto the sample (S). The modulation arm uses a combination of a half-wave plate and polarizer for a first attenuation, a second attenuation is achieved by a variable neutral density filter wheel. The excitation arm is attenuated by a series of neutral density filters. A schematic of the experiment can be found in the supplementary information (Fig. S1).
 
\paragraph*{Deactivation measurements.}
For the deactivation measurements, we removed the vortex phase plate from the modulation arm. A telescope focuses the modulation pulse so that it has an FWHM of 200 µm while the excitation NIR pulse has an FWHM of 40 µm. A tube lens focuses the back-reflected harmonics into a fiber spectrometer (Ocean Insight, QE Pro), that resolves the spectral composition of the harmonics.

\paragraph*{PSF imaging.}
A microscope objective (Mitutoyo Plan Apo NIR B) collimates the light from the sample with a magnification of 20x and an NA of 0.4. Subsequently, a tube lens with a focal length of 20 cm images them onto a camera. A pair of Bandpass filters blocked the donut pulse for PSF imaging. They have a central wavelength of 600 nm and a FWHM of 40 nm. The donut pulse was attenuated by an ND filter with an OD of 4.0 to avoid permanent damage. The images were taken \\
We evaluated the FWHM of the PSF by binarizing the images at a relative intensity of 0.5. Then an ellipse was fitted to the mask created that way. The FWHM represents the average radius of the fitted ellipse.

\paragraph*{Microscopy pictures.}
For the microscopy images both arms had to be attenuated by an effective optical density of 5. Otherwise, the high pulse energy of the laser could damage the thin-film surface. An 800 nm pulse deactivated the third harmonic response. To make scanning images we used a lens with a focal length of 5 cm and detected the reflected third harmonic emission. A silicon-based analog APD (Thorlabs) produces the signal out of the third harmonic generation. A preamplifier amplified the voltage output of the APD before the pulsed signal was integrated using a boxcar integrator. The Voltage output was then digitized by a data acquisition card. A piezo stage moved the sample through the focused pulse from spot to spot. \\
The images have a pixel size of 1 µm in both directions and are 51 x 51 pixels large. The images were scanned line by line. For the THG microscopy at each pixel 25 laser pulses were measured and each line was repeated 3 times, while for HADES each line was repeated 12 times with 25 laser pulses per pixel. The microscopy pictures displayed are the averaged values of each measurement, normalized to the maximum and minimum values. For the profile generation, the images were smoothed using a median filter with a pixel radius of 1.5. The profiles were generated with the profile plotter of ImageJ. The profiles were again normalized and interpolated by a spline algorithm for measurements of the FWHM of the features and measurements of the peak separation.

\section*{Acknowledgments}
\paragraph*{Funding:}
This work was conducted at the Advanced Research Center for Nanolithography, a public-private partnership between the University of Amsterdam (UvA), Vrije Universiteit Amsterdam (VU), Rijksuniversiteit Groningen (RUG), the Netherlands Organization for Scientific Research (NWO), and the semiconductor equipment manufacturer ASML and was (partly) financed by ‘Toeslag voor Topconsortia voor Kennis en Innovatie (TKI)’ from the Dutch Ministry of Economic Affairs and Climate Policy.
This manuscript is part of a project that has received funding from the European Research Council (ERC) under the European Union’s Horizon Europe research and innovation programme (grant agreement no. 101041819, ERC Starting Grant ANACONDA). 
Z.N., L.G., S.W. and P.M.K. acknowledge support from the Open Technology Programme (OTP) by NWO, grant no. 18703. 
The project is also part of the VIDI research programme HIMALAYA with project number VI.Vidi.223.133 financed by NWO.
\paragraph*{Author contribution:}
K.M., L.G., S.W., and P.M.K. envisioned and conceptualized the experiments. K.M. and P.M.K. carried out a preliminary experiment, and K.M., M.L.S.G., and P.E. carried out the main experiments. K.M. and Z.N. analyzed the data. L.G. and Z.N. provided and characterized the sample. P.M.K. supervised the work. K.M. and P.M.K. wrote a first version of the manuscript that was finalized by all authors.
\paragraph*{Competing interest:}
The authors declare that they have no competing interests. 
\paragraph*{Data availability:}
All data needed to evaluate the conclusions in the paper are present in the paper and/or the Supplementary Materials.

\bibliography{scibib}

\begin{thebibliography}{10}

\bibitem{Weisenburger2015}
S.~Weisenburger, V.~Sandoghdar, Light microscopy: an ongoing contemporary
  revolution.
\newblock {\it Contemporary Physics\/} {\bf 56}, 123-143 (2015).

\bibitem{Mertz2019}
J.~Mertz, {\it Introduction to optical microscopy\/} (Cambridge University
  Press, 2019).

\bibitem{Duncan1982}
M.~D. Duncan, J.~Reintjes, T.~J. Manuccia, Scanning coherent anti-stokes raman
  microscope.
\newblock {\it Optics Letters\/} {\bf 7}, 350--352 (1982).

\bibitem{Debarre2006}
D.~D{\'{e}}barre, W.~Supatto, A.~M. Pena, A.~Fabre, T.~Tordjmann, L.~Combettes,
  M.~C. Schanne-Klein, E.~Beaurepaire, {Imaging lipid bodies in cells and
  tissues using third-harmonic generation microscopy}.
\newblock {\it Nature Methods\/} {\bf 3}, 47--53 (2006).

\bibitem{Witte2011}
S.~Witte, A.~Negrean, J.~C. Lodder, C.~P. {De Kock}, G.~T. Silva, H.~D.
  Mansvelder, M.~L. Groot, {Label-free live brain imaging and targeted patching
  with third-harmonic generation microscopy}.
\newblock {\it Proceedings of the National Academy of Sciences of the United
  States of America\/} {\bf 108}, 5970--5975 (2011).

\bibitem{Kallioniemi2020}
L.~Kallioniemi, S.~Annurakshita, G.~Bautista, {Third-harmonic generation
  microscopy of undeveloped photopolymerized structures}.
\newblock {\it OSA Continuum\/} {\bf 3}, 2961 (2020).

\bibitem{Zhou2020}
L.~Zhou, H.~Fu, T.~Lv, C.~Wang, H.~Gao, D.~Li, L.~Deng, W.~Xiong, {Nonlinear
  optical characterization of 2d materials}.
\newblock {\it Nanomaterials\/} {\bf 10}, 1--38 (2020).

\bibitem{Yildirim2015}
M.~Yildirim, N.~Durr, A.~Ben-Yakar, {Tripling the maximum imaging depth with
  third-harmonic generation microscopy}.
\newblock {\it Journal of Biomedical Optics\/} {\bf 20}, 096013 (2015).

\bibitem{Murakami2022}
Y.~Murakami, M.~Masaki, S.~Miyazaki, R.~Oketani, Y.~Hayashi, M.~Yanagisawa,
  S.~Honjoh, H.~Kano, {Spectroscopic second and third harmonic generation
  microscopy using a femtosecond laser source in the third near-infrared
  (NIR-III) optical window}.
\newblock {\it Biomedical Optics Express\/} {\bf 13}, 694 (2022).

\bibitem{Shao2011}
L.~Shao, P.~Kner, E.~H. Rego, M.~G. Gustafsson, Super-resolution 3d microscopy
  of live whole cells using structured illumination.
\newblock {\it Nature methods\/} {\bf 8}, 1044--1046 (2011).

\bibitem{mcpherson87a}
A.~McPherson, G.~Gibson, H.~Jara, U.~Johann, T.~S. Luk, I.~McIntyre, K.~Boyer,
  C.~K. Rhodes, Studies of multiphoton production of vacuum-ultraviolet
  radiation in the rare gases.
\newblock {\it JOSA B\/} {\bf 4}, 595--601 (1987).

\bibitem{Ferray1988}
M.~Ferray, A.~L'Huillier, X.~F. Li, L.~A. Lompre, G.~Mainfray, C.~Manus,
  Multiple-harmonic conversion of 1064 nm radiation in rare gases.
\newblock {\it Journal of Physics B: Atomic, Molecular and Optical Physics\/}
  {\bf 21}, L31 (1988).

\bibitem{Ghimire2011}
S.~Ghimire, A.~D. Dichiara, E.~Sistrunk, P.~Agostini, L.~F. Dimauro, D.~A.
  Reis, {Observation of high-order harmonic generation in a bulk crystal}.
\newblock {\it Nature Physics\/} {\bf 7}, 138--141 (2011).

\bibitem{juergens19a}
P.~J{\"u}rgens, M.~Vrakking, A.~Husakou, R.~Stoian, A.~Mermillod-Blondin,
  Plasma formation and relaxation dynamics in fused silica driven by
  femtosecond short-wavelength infrared laser pulses.
\newblock {\it Applied Physics Letters\/} {\bf 115} (2019).

\bibitem{golde08a}
D.~Golde, T.~Meier, S.~W. Koch, High harmonics generated in semiconductor
  nanostructures by the coupled dynamics of optical inter-and intraband
  excitations.
\newblock {\it Physical Review B\/} {\bf 77}, 075330 (2008).

\bibitem{schubert14a}
O.~Schubert, M.~Hohenleutner, F.~Langer, B.~Urbanek, C.~Lange, U.~Huttner,
  D.~Golde, T.~Meier, M.~Kira, S.~W. Koch, {\it et~al.\/}, Sub-cycle control of
  terahertz high-harmonic generation by dynamical bloch oscillations.
\newblock {\it Nature photonics\/} {\bf 8}, 119--123 (2014).

\bibitem{shiner11a}
A.~Shiner, B.~Schmidt, C.~Trallero-Herrero, H.~J. W{\"o}rner, S.~Patchkovskii,
  P.~B. Corkum, J.~Kieffer, F.~L{\'e}gar{\'e}, D.~Villeneuve, Probing
  collective multi-electron dynamics in xenon with high-harmonic spectroscopy.
\newblock {\it Nature Physics\/} {\bf 7}, 464--467 (2011).

\bibitem{Smirnova2009}
O.~Smirnova, Y.~Mairesse, S.~Patchkovskii, N.~Dudovich, D.~Villeneuve,
  P.~Corkum, M.~Y. Ivanov, {High harmonic interferometry of multi-electron
  dynamics in molecules}.
\newblock {\it Nature\/} {\bf 460}, 972--977 (2009).

\bibitem{kraus15b}
P.~M. Kraus, B.~Mignolet, D.~Baykusheva, A.~Rupenyan, L.~Horn{\`y}, E.~F.
  Penka, G.~Grassi, O.~I. Tolstikhin, J.~Schneider, F.~Jensen, {\it et~al.\/},
  Measurement and laser control of attosecond charge migration in ionized
  iodoacetylene.
\newblock {\it Science\/} {\bf 350}, 790--795 (2015).

\bibitem{ghimire19a}
S.~Ghimire, D.~A. Reis, High-harmonic generation from solids.
\newblock {\it Nature physics\/} {\bf 15}, 10--16 (2019).

\bibitem{Kraus2013}
P.~M. Kraus, S.~B. Zhang, A.~Gijsbertsen, R.~R. Lucchese, N.~Rohringer, H.~J.
  W{\"{o}}rner, {High-harmonic probing of electronic coherence in dynamically
  aligned molecules}.
\newblock {\it Physical Review Letters\/} {\bf 111}, 1--5 (2013).

\bibitem{Wang2017}
Z.~Wang, H.~Park, Y.~H. Lai, J.~Xu, C.~I. Blaga, F.~Yang, P.~Agostini, L.~F.
  DiMauro, {The roles of photo-carrier doping and driving wavelength in high
  harmonic generation from a semiconductor}.
\newblock {\it Nature Communications\/} {\bf 8}, 1--7 (2017).

\bibitem{Heide2022}
C.~Heide, Y.~Kobayashi, A.~C. Johnson, F.~Liu, T.~F. Heinz, D.~A. Reis,
  S.~Ghimire, {Probing electron-hole coherence in strongly driven 2D materials
  using high-harmonic generation}.
\newblock {\it Optica\/} {\bf 9}, 512 (2022).

\bibitem{Wang2022}
Y.~Wang, F.~Iyikanat, X.~Bai, X.~Hu, S.~Das, Y.~Dai, Y.~Zhang, L.~Du, S.~Li,
  H.~Lipsanen, F.~J. {Garc{\'{i}}a De Abajo}, Z.~Sun, {Optical Control of
  High-Harmonic Generation at the Atomic Thickness}.
\newblock {\it Nano Letters\/} {\bf 22}, 8455--8462 (2022).

\bibitem{Bionta2021}
M.~R. Bionta, E.~Haddad, A.~Leblanc, V.~Gruson, P.~Lassonde, H.~Ibrahim,
  J.~Chaillou, N.~{\'{E}}mond, M.~R. Otto, {\'{A}}.~Jim{\'{e}}nez-Gal{\'{a}}n,
  R.~E. Silva, M.~Ivanov, B.~J. Siwick, M.~Chaker, F.~L{\'{e}}gar{\'{e}},
  {Tracking ultrafast solid-state dynamics using high harmonic spectroscopy}.
\newblock {\it Physical Review Research\/} {\bf 3}, 1--12 (2021).

\bibitem{Cheng2020}
Y.~Cheng, H.~Hong, H.~Zhao, C.~Wu, Y.~Pan, C.~Liu, Y.~Zuo, Z.~Zhang, J.~Xie,
  J.~Wang, D.~Yu, Y.~Ye, S.~Meng, K.~Liu, {Ultrafast Optical Modulation of
  Harmonic Generation in Two-Dimensional Materials}.
\newblock {\it Nano Letters\/}  (2020).

\bibitem{Geest2023}
M.~L.~S. van~der Geest, J.~J. de~Boer, K.~Murzyn, P.~Jürgens, B.~Ehrler, P.~M.
  Kraus, Transient high-harmonic spectroscopy in an inorganic–organic lead
  halide perovskite.
\newblock {\it The Journal of Physical Chemistry Letters\/} {\bf 14},
  10810-10818 (2023). PMID: 38015825.

\bibitem{VanEssen2024}
P.~J. van Essen, Z.~Nie, B.~de~Keijzer, P.~M. Kraus, Towards complete
  all-optical emission control of high-harmonic generation from solids (2024).

\bibitem{Brown2022}
G.~G. Brown, {\'{A}}.~Jim{\'{e}}nez-Gal{\'{a}}n, R.~E.~F. Silva, M.~Ivanov, {A
  Real-Space Perspective on Dephasing in Solid-State High Harmonic Generation}.
\newblock {\it arXiv\/} pp. 1--6 (2022).

\bibitem{Wang2023}
Y.~Wang, Y.~Liu, P.~Jiang, Y.~Gao, H.~Yang, L.~Y. Peng, Q.~Gong, C.~Wu,
  {Optical switch of electron-hole and electron-electron collisions in
  semiconductors}.
\newblock {\it Physical Review B\/} {\bf 107}, 1--5 (2023).

\bibitem{Weber2022}
M.~Weber, H.~von~der Emde, M.~Leutenegger, P.~Gunkel, S.~Sambandan, T.~A. Khan,
  J.~Keller-Findeisen, V.~C. Cordes, S.~W. Hell, {MINSTED nanoscopy enters the
  {\AA}ngstr{\"{o}}m localization range}.
\newblock {\it Nature Biotechnology\/} {\bf 41} (2023).

\bibitem{Betzig1995}
E.~Betzig, Proposed method for molecular optical imaging.
\newblock {\it Opt. Lett.\/} {\bf 20}, 237--239 (1995).

\bibitem{Betzig2006}
E.~Betzig, G.~H. Patterson, R.~Sougrat, O.~W. Lindwasser, S.~Olenych, J.~S.
  Bonifacino, M.~W. Davidson, J.~Lippincott-Schwartz, H.~F. Hess, {Imaging
  intracellular fluorescent proteins at nanometer resolution}.
\newblock {\it Science\/} {\bf 313}, 1642--1645 (2006).

\bibitem{Rust2006}
M.~J. Rust, M.~Bates, X.~Zhuang, {Sub-diffraction-limit imaging by stochastic
  optical reconstruction microscopy (STORM)}.
\newblock {\it Nature Methods\/} {\bf 3}, 793--795 (2006).

\bibitem{Sharonov2006}
A.~Sharonov, R.~M. Hochstrasser, {Wide-field subdiffraction imaging by
  accumulated binding of diffusing probes}.
\newblock {\it Proceedings of the National Academy of Sciences of the United
  States of America\/} {\bf 103}, 18911--18916 (2006).

\bibitem{Hell1994}
S.~W. Hell, J.~Wichmann, {Breaking the diffraction resolution limit by
  stimulated emission: stimulated-emission-depletion fluorescence microscopy}.
\newblock {\it Optics Letters\/} {\bf 19}, 780 (1994).

\bibitem{Klar2000}
T.~A. Klar, S.~Jakobs, M.~Dyba, A.~Egner, S.~W. Hell, {Fluorescence microscopy
  with diffraction resolution barrier broken by stimulated emission}.
\newblock {\it Proceedings of the National Academy of Sciences of the United
  States of America\/} {\bf 97}, 8206--8210 (2000).

\bibitem{Hell1995}
S.~W. Hell, M.~Kroug, Ground-state-depletion fluorscence microscopy: A concept
  for breaking the diffraction resolution limit.
\newblock {\it Applied Physics B\/} {\bf 60}, 495--497 (1995).

\bibitem{Gustafsson2005}
M.~G. Gustafsson, {Nonlinear structured-illumination microscopy: Wide-field
  fluorescence imaging with theoretically unlimited resolution}.
\newblock {\it Proceedings of the National Academy of Sciences of the United
  States of America\/} {\bf 102}, 13081--13086 (2005).

\bibitem{Nie2023}
Z.~Nie, L.~Guery, E.~B. Molinero, P.~Juergens, T.~J. van~den Hooven, Y.~Wang,
  A.~Jimenez~Galan, P.~C.~M. Planken, R.~E.~F. Silva, P.~M. Kraus, Following
  the nonthermal phase transition in niobium dioxide by time-resolved harmonic
  spectroscopy.
\newblock {\it Phys. Rev. Lett.\/} {\bf 131}, 243201 (2023).

\bibitem{Pinhas2018}
H.~Pinhas, O.~Wagner, Y.~Danan, M.~Danino, Z.~Zalevsky, M.~Sinvani, {Plasma
  dispersion effect based super-resolved imaging in silicon}.
\newblock {\it Optics Express\/} {\bf 26}, 25370 (2018).

\bibitem{Wang2013}
P.~Wang, M.~N. Slipchenko, J.~Mitchell, C.~Yang, E.~O. Potma, X.~Xu, J.~X.
  Cheng, {Far-field imaging of non-fluorescent species with subdiffraction
  resolution}.
\newblock {\it Nature Photonics\/} {\bf 7}, 449--453 (2013).

\bibitem{Kanou2017}
H.~Kanou, Y.~Iketaki, Super-resolution microscope (Japan Patent JP2018120006A,
  Jan. 2017).

\bibitem{penwell2017}
S.~B. Penwell, L.~D. Ginsberg, R.~Noriega, N.~S. Ginsberg, Resolving ultrafast
  exciton migration in organic solids at the nanoscale.
\newblock {\it Nature Materials\/} {\bf 16}, 1136--1141 (2017).

\bibitem{yang2011}
Z.~Yang, C.~Ko, S.~Ramanathan, Oxide electronics utilizing ultrafast
  metal-insulator transitions.
\newblock {\it Annual Review of Materials Research\/} {\bf 41}, 337--367
  (2011).

\bibitem{kumar2020}
S.~Kumar, R.~S. Williams, Z.~Wang, Third-order nanocircuit elements for
  neuromorphic engineering.
\newblock {\it Nature\/} {\bf 585}, 518--523 (2020).

\bibitem{abbing2020}
S.~R. Abbing, F.~Campi, F.~S. Sajjadian, N.~Lin, P.~Smorenburg, P.~M. Kraus,
  Divergence control of high-harmonic generation.
\newblock {\it Physical Review Applied\/} {\bf 13}, 054029 (2020).

\bibitem{roscam2021}
S.~D. Roscam~Abbing, F.~Campi, A.~Zeltsi, P.~Smorenburg, P.~M. Kraus,
  Divergence and efficiency optimization in polarization-controlled two-color
  high-harmonic generation.
\newblock {\it Scientific Reports\/} {\bf 11}, 24253 (2021).

\bibitem{Negro2011}
M.~Negro, C.~Vozzi, K.~Kovacs, C.~Altucci, R.~Velotta, F.~Frassetto,
  L.~Poletto, P.~Villoresi, S.~de~Silvestri, V.~Tosa, S.~Stagira, {Gating of
  high-order harmonics generated by incommensurate two-color mid-IR laser
  pulses}.
\newblock {\it Laser Physics Letters\/} {\bf 8}, 875--879 (2011).

\bibitem{becker88a}
P.~Becker, H.~Fragnito, C.~B. Cruz, R.~Fork, J.~Cunningham, J.~Henry, C.~Shank,
  Femtosecond photon echoes from band-to-band transitions in gaas.
\newblock {\it Physical review letters\/} {\bf 61}, 1647 (1988).

\bibitem{Westphal2005}
V.~Westphal, S.~W. Hell, {Nanoscale resolution in the focal plane of an optical
  microscope}.
\newblock {\it Physical Review Letters\/} {\bf 94}, 1--4 (2005).

\bibitem{RoscamAbbing2022}
S.~D. {Roscam Abbing}, R.~Kolkowski, Z.~Y. Zhang, F.~Campi, L.~L{\"{o}}tgering,
  A.~F. Koenderink, P.~M. Kraus, {Extreme-Ultraviolet Shaping and Imaging by
  High-Harmonic Generation from Nanostructured Silica}.
\newblock {\it Physical Review Letters\/} {\bf 128}, 223902 (2022).

\bibitem{Beeker2011}
W.~P. Beeker, C.~J. Lee, K.~J. Boller, P.~Gro{\ss}, C.~Cleff, C.~Fallnich,
  H.~L. Offerhaus, J.~L. Herek, {A theoretical investigation of
  super-resolution CARS imaging via coherent and incoherent saturation of
  transitions}.
\newblock {\it Journal of Raman Spectroscopy\/} {\bf 42}, 1854--1858 (2011).

\bibitem{Zhang2010}
S.~Zhang, J.~Shi, H.~Zhang, T.~Jia, Z.~Wang, Z.~Sun, {Precise control of
  state-selective excitation in stimulated Raman scattering}.
\newblock {\it Physical Review A - Atomic, Molecular, and Optical Physics\/}
  {\bf 82}, 4--7 (2010).

\bibitem{Sekiguchi2008}
K.~Sekiguchi, S.~Yamaguchi, T.~Tahara, {Femtosecond time-resolved electronic
  sum-frequency generation spectroscopy: A new method to investigate ultrafast
  dynamics at liquid interfaces}.
\newblock {\it Journal of Chemical Physics\/} {\bf 128} (2008).

\bibitem{wang2015}
Y.~Wang, R.~B. Comes, S.~Kittiwatanakul, S.~A. Wolf, J.~Lu, Epitaxial niobium
  dioxide thin films by reactive-biased target ion beam deposition.
\newblock {\it Journal of Vacuum Science \& Technology A\/} {\bf 33} (2015).

\end{thebibliography}

\bibliographystyle{ScienceAdvances}

\newpage
\begin{figure}[htb!]
\includegraphics[scale=1]{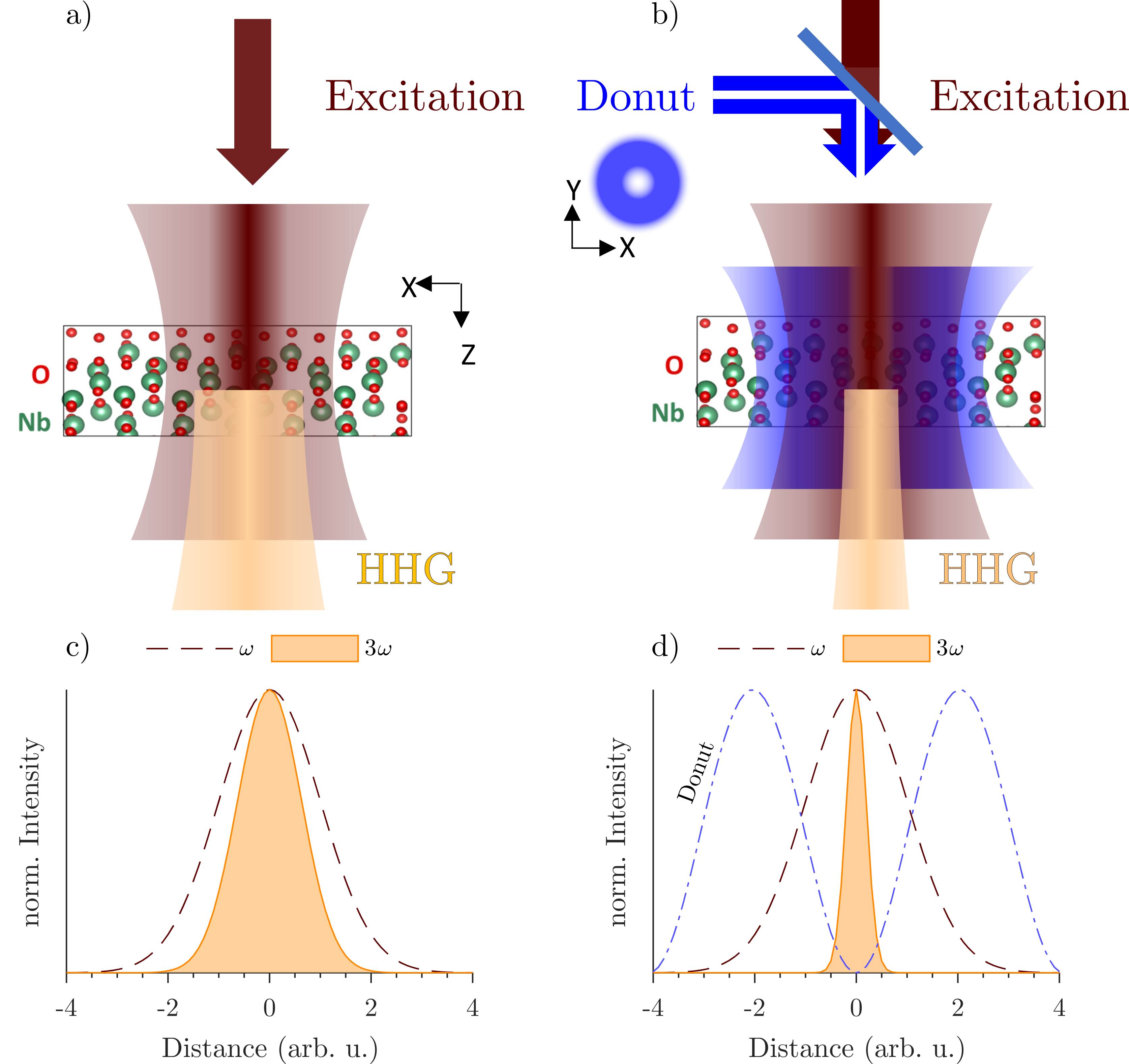}
\caption{\label{fig:concept} \textbf{Principle of HADES.} Schematic of the illumination scheme (a,c) and normalized intensities in the focal spot (b,d) for high harmonic generation (HHG) microscopy (a,c) and for harmonic deactivation microscopy (HADES)(b,d). In an HHG microscope a fundamental laser pulse at central frequency $\omega$ excites HHG. Compared to this an additional donut-shaped pulse deactivates high harmonic emission outside of the center. The resulting spot size which effectively can be observed is reduced significantly.}
\end{figure}

\newpage
\begin{figure}[htp!]
\includegraphics[scale=1]{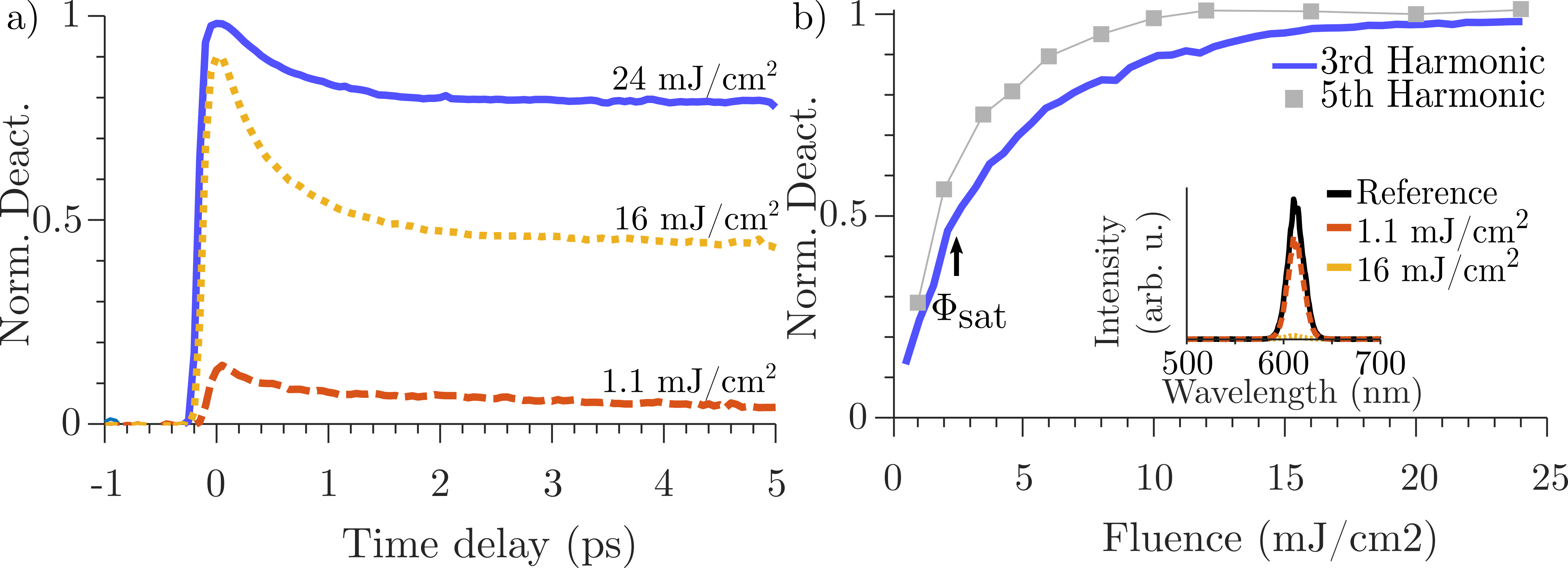}
\caption{\label{fig:experiment_setup}\textbf{Deactivation mechanism.} a) Time-dependent normalized deactivation of the third harmonic response plotted at rising fluences of the modulation pulse. At negative times the NIR excitation pulse arrives first. The strongest deactivation is always seen at the temporal overlap of the two pulses. b) Fluence-dependent deactivation of the third harmonic (blue) and fifth harmonic (grey, taken from \cite{Nie2023} normalized to the reference harmonic emission without modulation pulse. The saturation fluence $\phi_{sat}$ is defined as the point of 50 \% normalized deactivation. The fifth harmonic shows steeper deactivation and potentially better resolution improvement. The inset shows the third harmonic spectra of the reference (blue) and the spectra with a modulation fluence of 1 mJ/cm$^2$ (orange) and a modulation fluence of 16 mJ/cm$^2$ (yellow).}
\end{figure}

\newpage
\begin{figure}[htp!]
\begin{center}
\includegraphics[scale=1]{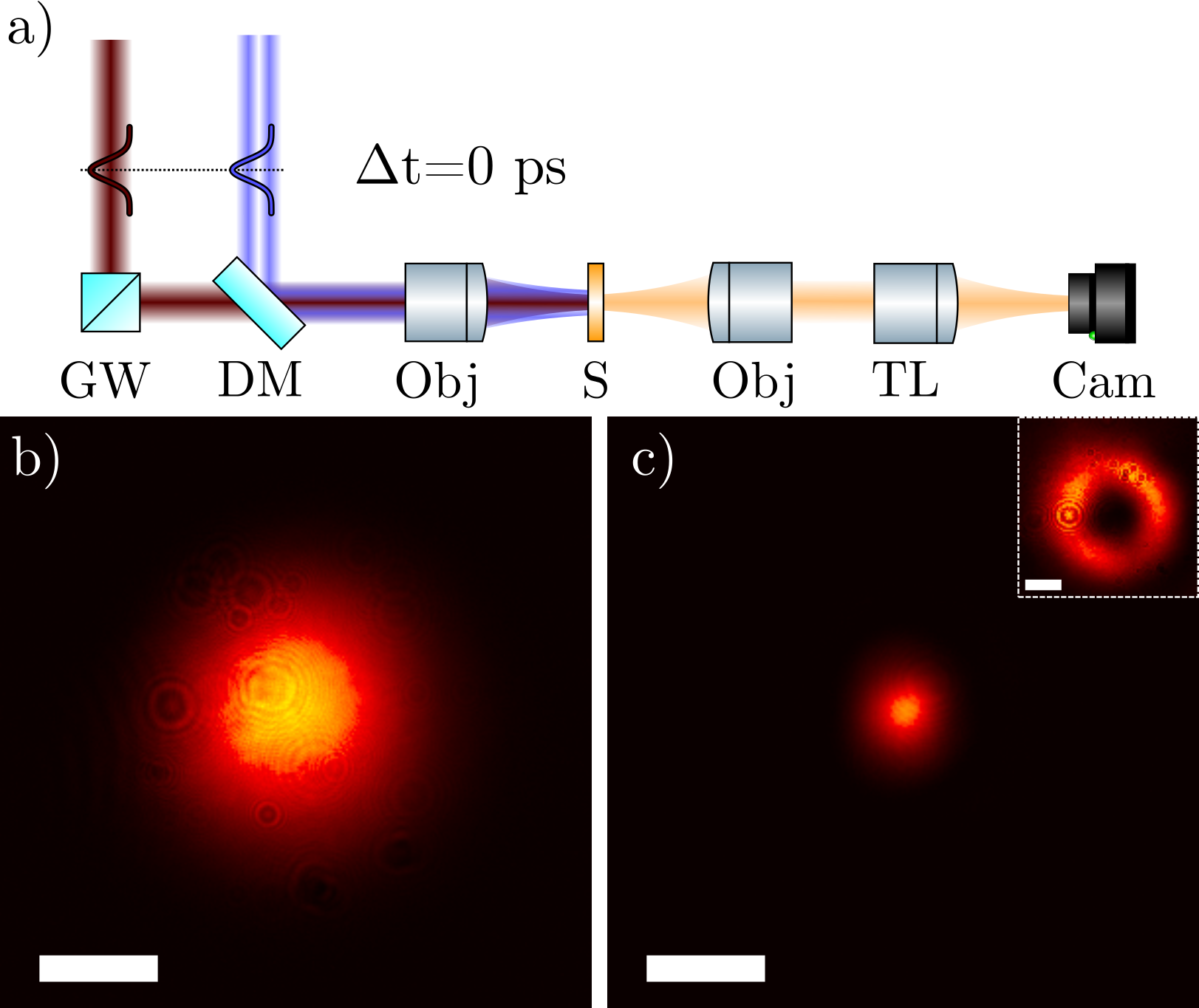}
\end{center}
\caption{\textbf{Point-spread function reduction with donut pulse.} The excitation pulse (red) and donut pulse (blue) get combined and focused onto the sample by a lens with a 30 cm focal length. A home-built microscope images the point-spread function (PSF) of the transmitted harmonics (a). PSF on the sample without (b) and with (c) the OAM pulse, the inset shows the deactivation pulse. The bar is 30 µm, which is the resolution limit of the focusing objective at the THG wavelength. GW, glass wedge; DM, dichroic mirror; Obj, Objective lens; S, sample; TL, tube lens; Cam, camera. \label{fig:PSF}}
\end{figure}

\newpage
\begin{figure}[htp!]
\begin{center}
\includegraphics[scale=1]{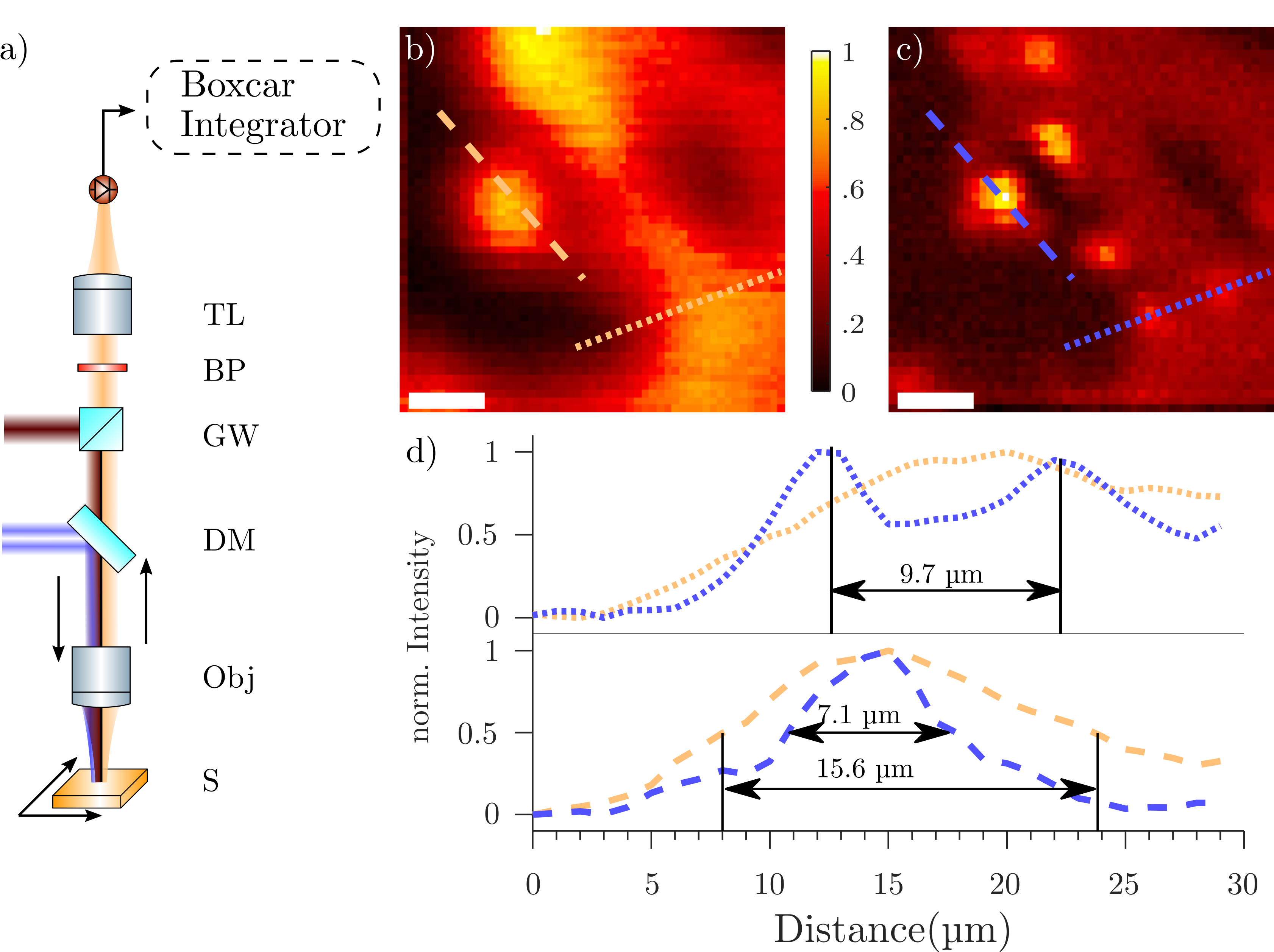}
\end{center}
\caption{\textbf{Resolution improvement through HADES.} A NIR (red) and donut (blue) pulse are focused onto the sample (a). The back-reflected third-harmonics (orange) are detected by an APD. Scanning the sample through the focus spot generates the image. Blocking the donut pulse results in a third-harmonic generation (THG) image (b). Adding the donut pulse (c) gives rise to an image through harmonic deactivation microscopy (HADES). The color bar applies to both images and shows the normalized intensity. The scale bar is 10 µm. The lower panel (d) shows profiles along the dotted line and the dashed line in the above pictures. The profiles in blue are taken from the THG microscopy picture and in orange from the HADES picture. These profiles show a significant resolution improvement for HADES over THG microscopy. GW, glass wedge; DM, dichroic mirror; Obj, objective lens; S, sample; TL, tube lens; BP, bandpass filter; APD, avalanche photodiode. \label{fig:Microscopy}}
\end{figure} 

\newpage


\section*{Supplementary material}

# Supplementary Materials for

## Breaking Abbe's diffraction limit with harmonic deactivation microscopy

Kevin Murzyn et al.

*Corresponding author. Email: k.murzyn@arcnl.nl, p.kraus@arcnl.nl

**This PDF file includes:**

    Supplementary Text
    Figs. S1 to S5

**Supplementary Text**

Experiments

The experimental setup is a modified Mach-Zehnder interferometer (Fig. S1). The arm in the vertical direction is the generating arm and in the horizontal direction is the arm of the deactivation pulses. In the generating arm, an optical parametric amplifier (OPA) converts the wavelength of the fundamental pulse from 800 nm to 1800 nm (NIR pulse). In the deactivation arm, a vortex phase plate is inserted to generate the donut pulse. Three different ways enable the detection of emitted harmonics. A wide field microscope images the point-spread function (PSF). A UV/Vis spectrometer resolves the spectrum in reflection. An avalanche photodiode detects individual pulses for raster scanning imaging.

Numerical Simulations

The numerical simulations were conducted by assuming a Gaussian intensity distribution of arbitrary size for the third-harmonic generation (THG, orange line in Fig. S2). The deactivation curve from Fig. 2b was then interpolated onto a cos²-function with its minimum at the maximum of THG and the maximum of the cos² at the full-width half max (FWHM) of the THG. The two functions were multiplied to get the harmonic deactivation microscopy (HADES, blue line in Fig. S2). FWHMs of both methods were compared to get the relative resolution improvement.

Resolution limit of HADES

The formula for the resolution limit starts at Abbe's diffraction limit for the NIR excitation laser with wavelength $\lambda_0$:

$$d_0 = \frac{\lambda_0}{2 * \text{NA}}.$$

Where the $d_0$ is the minimum distance of two point sources. For focusing a laser pulse the effective NA can be calculated as:

$$\text{NA} = \frac{\pi * w_0}{4 * f},$$

With the pulse waist $w_0$ of the collimated pulse and the focus length f of the lens. We assume perturbative non-linear optics therefore the initial Gaussian-shaped pulse intensity of the n-th harmonic can be described as:

$$I_{\text{nth-HG}}(x, y) \sim I_{NIR}^n(x, y) = \exp\left(-\frac{x^2 + y^2}{(d_0/\sqrt{n})^2}\right),$$

Extracting the pulse waist leads to the resolution of the n-th harmonic as:

$$d = \frac{\lambda_0}{2 * \text{NA} * \sqrt{n}}.$$

The introduction of the resolution is based on the phenomenological observation that in the simulation the improvement will increase if the peak fluence increases. Therefore, we adapt the formula in (46) to be:

$$d_{HADES} = \frac{\lambda_0}{2 * \text{NA} * \sqrt{n}} * \frac{1}{\sqrt{1+\zeta}},$$

with the saturation level $\zeta$.

Convolution of HADES
The images obtained by HADES were convolved with a 2D gaussian distribution Fig. S3. The Gaussian distribution had to be made asymmetric and rotated. To minimize edge effect the convolution was carried out in accordance with the convolution theorem. The following formula was used for the gaussian distribution in the frequency space:

$$f(x,y) = \frac{1}{2\pi\sigma_x\sigma_y} \exp(-(ax^2 + bxy + cy^2)),$$

with

$$a = \frac{\cos^2(\theta)}{2\sigma_x} + \frac{\sin^2(\theta)}{2\sigma_y},$$

$$b = -\frac{\sin(2\theta)}{4\sigma_x} + \frac{\sin(2\theta)}{4\sigma_y},$$

$$c = \frac{\sin^2(\theta)}{2\sigma_x} + \frac{\cos^2(\theta)}{2\sigma_y}.$$

The parameters to obtain the resulting convolution are $\sigma_x = 0.08\ \mu m^{-1}$, $\sigma_y = 0.13\ \mu m^{-1}$ and $\theta = 45°$. The multiplication of the Fourier transformed picture with the gaussian was inverse Fourier transformed to get the convolution.


Light microscopy images
The bright-field microscopy (Fig. S4a) images were taken with an effective magnification of 80. The dark-field microscopy images were taken with an effective magnification of 50. The position was traced back through the absolute coordinates of the stage in the Mach-Zehnder interferometer in combination with markers on the sample. The Darkfield microscopy (Fig. S4b) images show a roughening of the surface of the sample.


Electron Microscopy images
In the electron microscopy images (Fig. S5), the nanostructure from the laser damage becomes visible. This nanostructure enhances the third harmonic generation of the thin film. The deep "L" shaped trench shows a complete ablation of the NbO2 film.

3

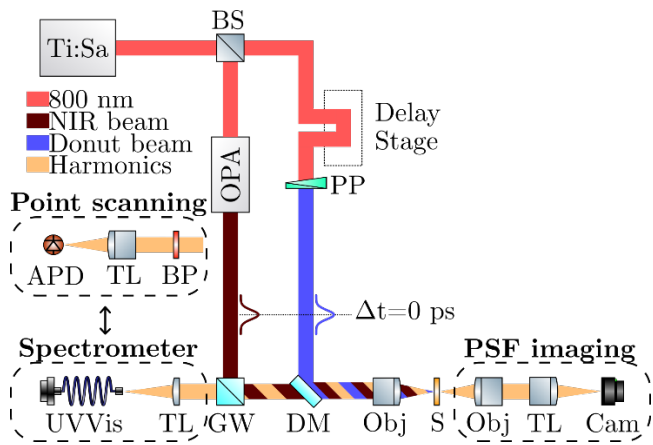

**Fig. S1. Experimental setup**

BS, pulse splitter; OPA, optical parametric amplifier; PP, vortex phase plate; GW, glass wedge; DM, dichroic mirror; Obj, objective lens; S, sample; TL, tube lens; Cam, camera; BP, bandpass filter; UVVis, Fiber spectrometer; APD, avalanche photodiode.

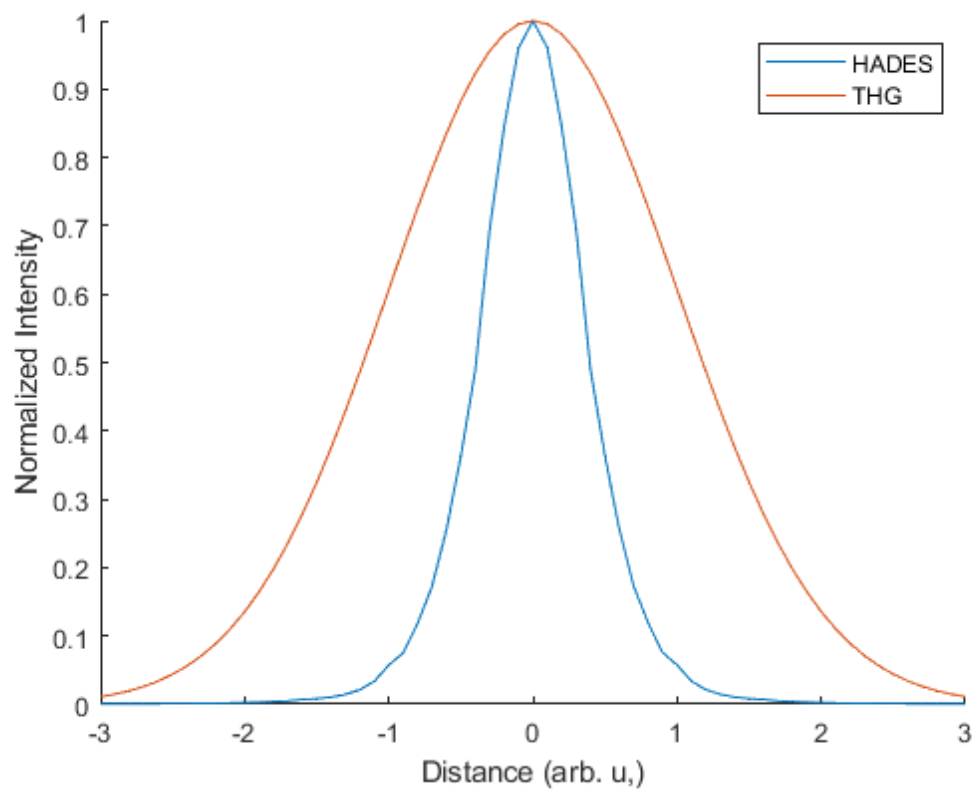

**Fig. S2. Numerical calculation of point-spread function**

Assumed Gaussian intensity distribution for third harmonic generation (orange). This is combined with a deactivation profile which is assumed to be of the form $\cos^2(x)$, where x is the distance. The deactivation is taken from the normalized deactivation in Fig. 2 of the main text. The peak fluence is assumed to be 25 mJ/cm². The resulting Gaussian peak of the harmonic deactivation microscopy (HADES) is smaller by a factor of three.

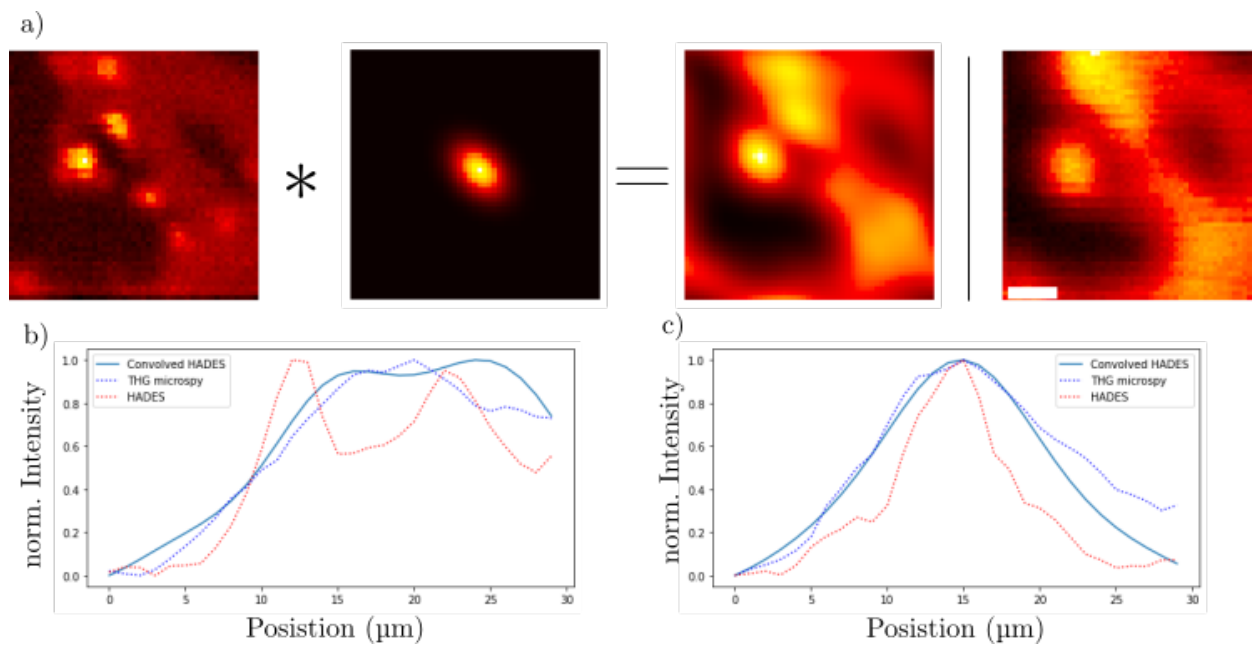

**Fig. S3. Convolutions of HADES images**

a) The whole image taken by HADES is convolved with a slightly asymmetric and rotated Gaussian distribution. The asymmetry takes into account imperfections in focusing of both driving and donut deactivation pulses. The resulting image resembles the image taken by third-harmonic generation (THG) microscopy (far right). b) Dotted profile as shown in Fig 4. of the main text. The solid line shows a convolution of the HADES profile with a Gaussian. Only on the edges of the profile lines does the convolution show bigger differences to the THG profile. c) Dashed profile as shown in Fig. 4 in the main text. The solid line shows the convolution of the HADES profile. At the right edge, the convolved lineout shows a larger difference to the THG profile due to the missing 2D information, which contains a blurred edge in the THG image but not in the HADES image.

6

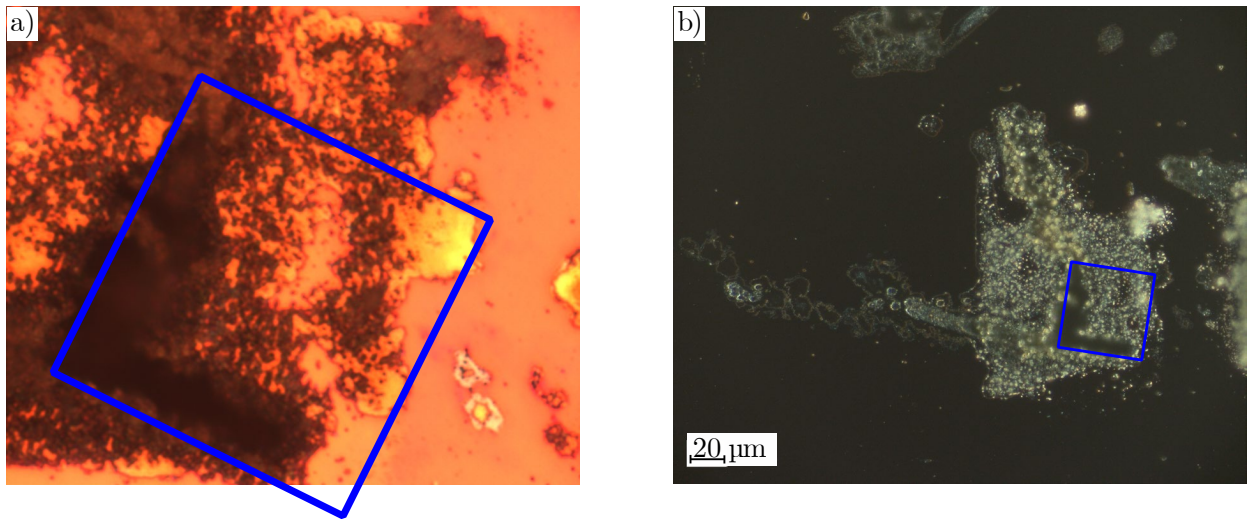

**Fig. S4. Full-field microscopy images**

Bright-field (a) and dark-field (b) microscopy images of the area images in Fig 4. In the dark field an increased surface roughness of the image can be seen. This indicates the thermal damage of the thin film. The "L"-shaped area is a complete ablation of the thin film.

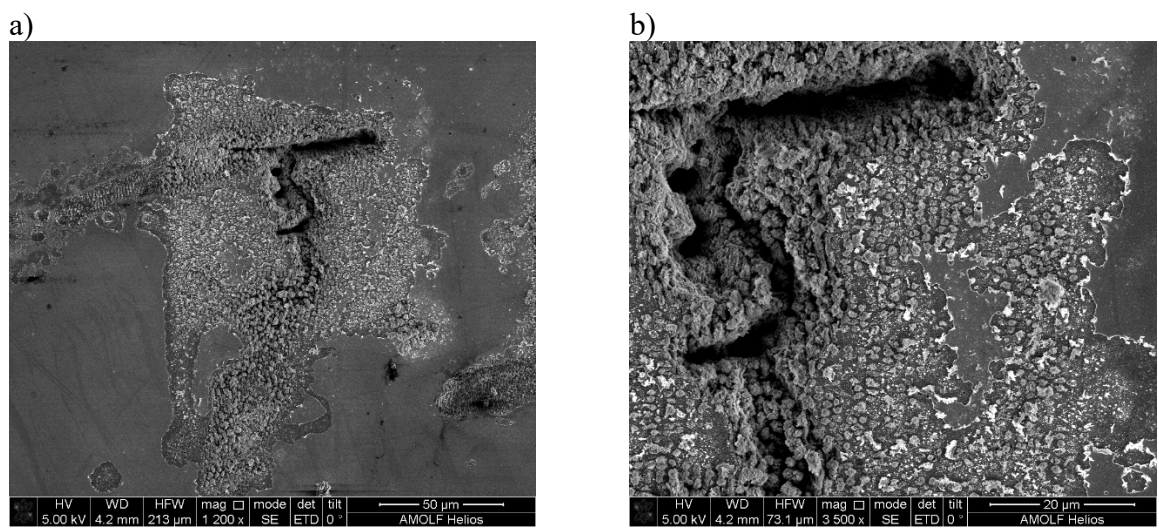

**Fig. S4. Scanning electron microscopy images**

a) Overview of the scanned area that got imaged with THG and HADES. Compared to the scanning and full-field images these images are flipped along the horizontal axis. b) Zoomed-in version of the scanned images shown before. The nanostructures on the surface are visible. It can be the reason for the enhancement of the THG at certain positions.


\end{document}